%% file: root.tex
\documentclass[letterpaper, 10 pt, conference]{ieeeconf}  

\IEEEoverridecommandlockouts                              

\usepackage{xcolor}
\usepackage{graphics} 
\usepackage{epsfig} 
\usepackage{amsmath} 
\usepackage{amssymb}  
\usepackage{svg}

\usepackage{todonotes}

\title{\LARGE \bf
FPGA-Based Neural Thrust Controller for UAVs 
}

\author{Sharif Azem, David Scheunert, Mengguang Li, Jonas Gehrunger,\\ Kai Cui, Christian Hochberger and Heinz Koeppl%
\thanks{**This work has been funded by the Distr@l "Digital Innovation and Technology Funding" of the Hessian Ministry for Digitalisation and Innovation (project no. 493 21\_0052\_2A).}
\thanks{The authors are with the Department of Electrical Engineering and Information Technology, Technische Universität Darmstadt, 64287 Darmstadt, Germany. {\tt\footnotesize\{sharif.azem, mengguang.li, kai.cui, heinz.koeppl\}@bcs.tu-darmstadt.de, \{scheunert, gehrunger, hochberger\}@rs.tu-darmstadt.de}}%
}

\begin{document}

\maketitle
\thispagestyle{empty}
\pagestyle{empty}

\input{main}
 
\addtolength{\textheight}{-12cm}   

\bibliographystyle{IEEEtran}
\bibliography{IEEEabrv, references}

\end{document}

%% file: main.tex
\begin{abstract}
The advent of unmanned aerial vehicles (UAVs) has improved a variety of fields by providing a versatile, cost-effective and accessible platform for implementing state-of-the-art algorithms. To accomplish a broader range of tasks, there is a growing need for enhanced on-board computing to cope with increasing complexity and dynamic environmental conditions. Recent advances have seen the application of Deep Neural Networks (DNNs), particularly in combination with Reinforcement Learning (RL), to improve the adaptability and performance of UAVs, especially in unknown environments. However, the computational requirements of DNNs pose a challenge to the limited computing resources available on many UAVs. This work explores the use of Field Programmable Gate Arrays (FPGAs) as a viable solution to this challenge, offering flexibility, high performance, energy and time efficiency. We propose a novel hardware board equipped with an Artix-7 FPGA for a popular open-source micro-UAV platform. We successfully validate its functionality by implementing an RL-based low-level controller using real-world experiments.
\end{abstract}

\section{INTRODUCTION}
As a platform for a relatively compact and cost-effective system capable of agile aerial performance, the development of UAVs has contributed to diverse fields such as agriculture, surveillance, disaster management and logistics. With their ability to access remote or hard-to-reach areas, UAVs have emerged as a tool for data collection \cite{8417673}, monitoring \cite{doi:10.1080/19475705.2017.1315619}, and rapid response missions \cite{drones3030059}. The application of UAVs often involves complex algorithms, such as collision avoidance \cite{9108245}, communication \cite{azem2022dynamic}, data processing \cite{mueller2015fusing}, low level control \cite{8967695} and indoor navigation \cite{9841385}. The increasing complexity of tasks assigned to UAVs requires advancements in onboard computation and processing capabilities to ensure a reliable and effective operation. On top of that, UAVs often need to react to sudden changes in the environment they operate in, necessitating adaptive algorithms that are applicable in such scenarios. 

Recently, DNNs have been used for the application of algorithms involving UAVs \cite{9812221}, \cite{batra2022decentralized}.  DNNs are often used in combination with RL, where an agent learns the optimal behavior through an interaction with its environment. The integration of DNNs on UAV-systems in the context of RL can enhance the performance of UAVs, allowing them to react to previously unseen situations and adapt to changing environments. However, Many UAVs have computers with limited computing power on board, making the implementation of DNNs infeasible. 

FPGAs offer a promising solution to the computational demands of deploying DNNs on UAVs. FPGAs provide flexibility in the implementation with the high performance and energy efficiency of hardware. Their re-configurable nature, as opposed to Application-Specific Integrated Circuits (ASICs), allows for an optimized implementation of different architectures of DNNs. Furthermore, the implementation of an algorithm can be designed to consume less power, compared to an implementation of the same algorithm on Graphics Processing Units (GPUs), which provides high computation power but also consumes more energy. 

In this work we present an implementation of DNNs on an expansion deck equipped with an FPGA for implementation of algorithms, in particular we implement DNNs with the deepsets architecture proposed in \cite{batra2022decentralized}. Furthermore, the expansion deck is based on the Lighthouse deck suggested in \cite{taffanel2021lighthouse} to allow decentralized position estimation. In this work, we refer to the deck as the Lighthouse-FPGA (LF) deck. The ability to calculate the next action and the own position on board contributes to the autonomous operation of the UAV, reducing the dependency on an external system. 

\begin{figure}[t]
  \centering
  \includegraphics[scale=.8]{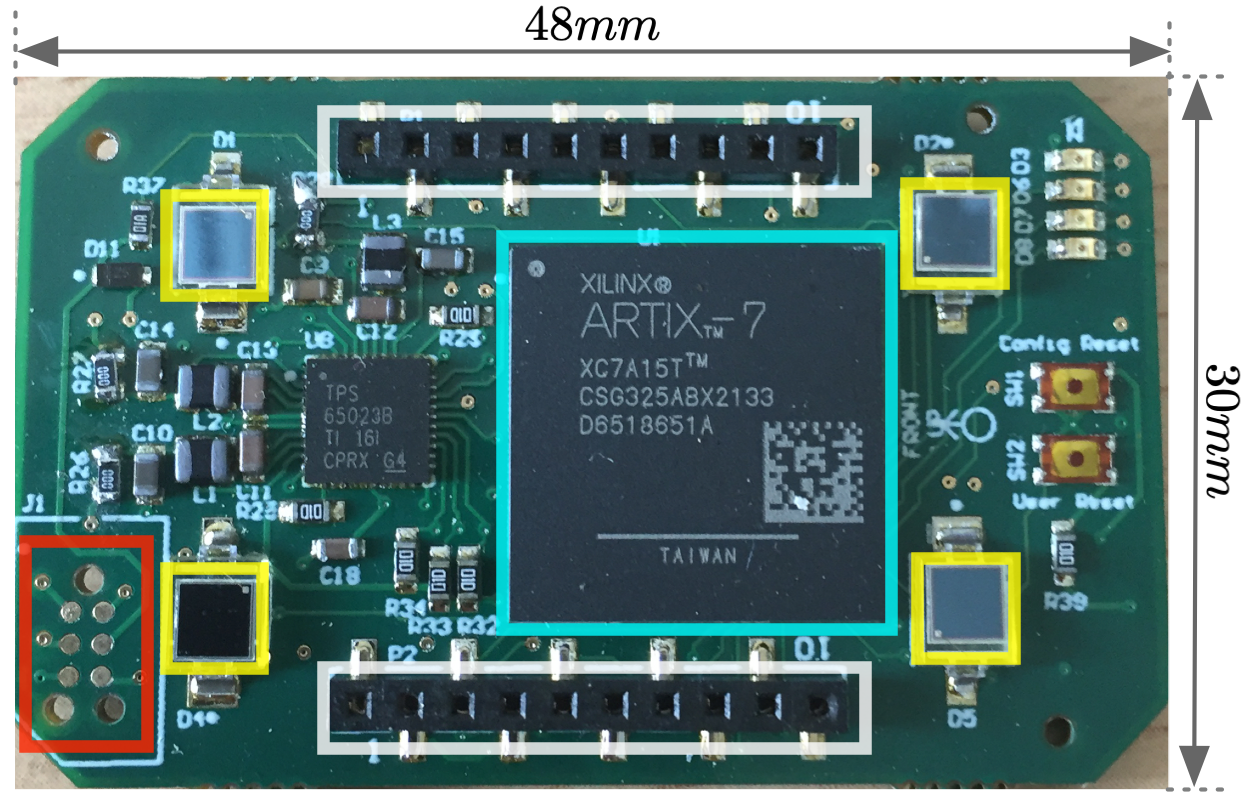}
  \caption{LF-deck with 4 IR receivers (marked in yellow), an Artix-7 FPGA XC7A15T (marked in light blue), one JTAG interface (marked in red) and an interface for connecting the deck with the Crazyflie (marked in white).}
  \label{frontpage}
\end{figure}

\section{SYSTEM}
The design of the LF expansion deck is divided into two parts with two different functionalities, both implemented on the FPGA of the deck. One functionality is responsible for receiving IR signals, and the other functionality is used for hardware acceleration, and consists of the MicroBlaze softcore processor\footnote{https://www.xilinx.com/products/design-tools/microblaze.html} and a hardware accelerator to run onboard computation faster. Furthermore, to reduce the resource utilization of the resulting hardware implementation, fixed point calculations are used. This leads to a simplified hardware design of the neural network as well as a much lower latency of the calculated output.

\subsection{Lighthouse-FPGA Expansion Deck}
\label{subsec:deck}

\begin{figure}[t]
\centering
\includegraphics[width=1.0\linewidth]{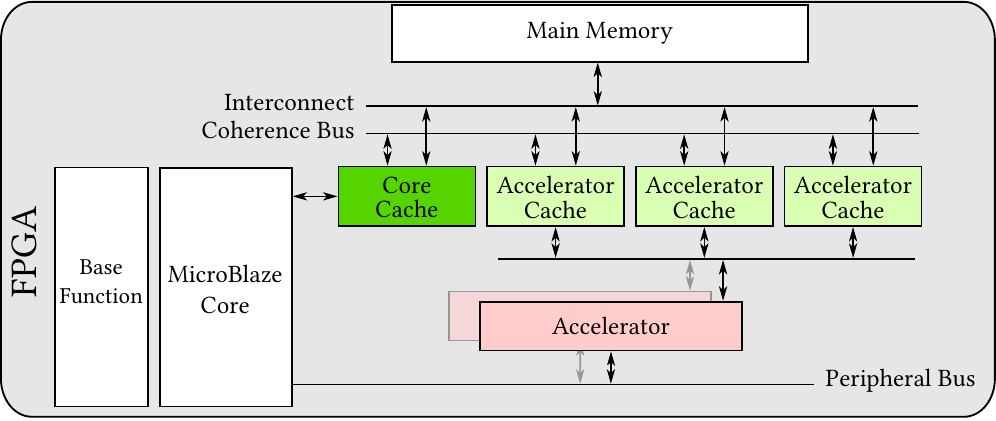}
\caption{Block diagram of the FPGA implementation. The base function refers to the functionality provided by the LF-deck suggested in \cite{taffanel2021lighthouse}. Additionally, a combination of a MicroBlaze softcore processor and a hardware accelerator is provided, where the hardware accelerator is created at compilation time using a tool called PIRANHA.}
\label{fig:block-diagram}
\end{figure}

In this study, we employ the Crazyflie nano quadrotor \cite{giernacki2017crazyflie}, equipped with the STM32F405 micro-controller (MCU) together with the Lighthouse deck.
 For this, we attach the deck on top of the Crazyflie through the Crazyflie's 20 pin interface (Figure \ref{frontpage} white marks). This interface is used both for power supply (from the Crazyflie's battery) and communication between the FPGA and the MCU.
The deck is equipped with four IR sensors (see Figure \ref{frontpage}, orange marks)  that can detect signals from Lighthouse base stations for on-board position estimation, and dedicated ICs  to process these signals. 
 The design and functionality are based on the Lighthouse deck proposed in \cite{taffanel2021lighthouse}. However, the LF-deck is equipped with an AMD/Xilinx Artix 7 FPGA XC7A15T, as in Figure \ref{frontpage}. The positioning module (in this work called also base function) is migrated to the LF-deck and communicates with the MCU through a Universal Asynchronous Receiver-Transmitter (UART) protocol.
 Furthermore, we use the MicroBlaze soft core processor alongside a hardware accelerator on the FPGA for additional calculations of computation intensive algorithms, such as the feed-forward calculation of DNNs.
Utilizing the I2C protocol bidirectionally, the MCU initially transmits the necessary information to the MicroBlaze for computation. Then, the computation is executed and the result is sent back to the MCU.
For programming the FPGA, we use a High-Level Synthesis (HLS) tool-chain that creates custom hardware accelerators, designed to identify parts of the implemented code to accelerate repetitive tasks, such as the execution of loops. The structure of the implementation within the FPGA is shown in Fig. \ref{fig:block-diagram}.
In order to flash the program onto the FPGA, The LF-deck is equipped with a JTAG interface (see Figure \ref{frontpage} red mark), which makes it possible to flash programs directly to the deck without using the Crazyflie.

\subsection{Hardware Accelerator}
\label{subsec:hw_accel}

The hardware accelerator is created through High-Level Synthesis (HLS). For this purpose, the tool \textit{Plugin for Intermediate Representation ANalysis and Hardware Acceleration} (PIRANHA) \cite{hempel2019generation} is used. It is a plugin for the GNU Compiler Collection (GCC), that operates on un-annotated source code and requires no manual integration of the accelerator into the source code by the developer. The selection of optimization strategies and hardware architecture of the entire System-on-Chip (SoC) realized on the FPGA is created using the SpartanMC system builder \cite{jconfig}.

\begin{figure}[h!]
\centering
\includegraphics[width=1\linewidth]{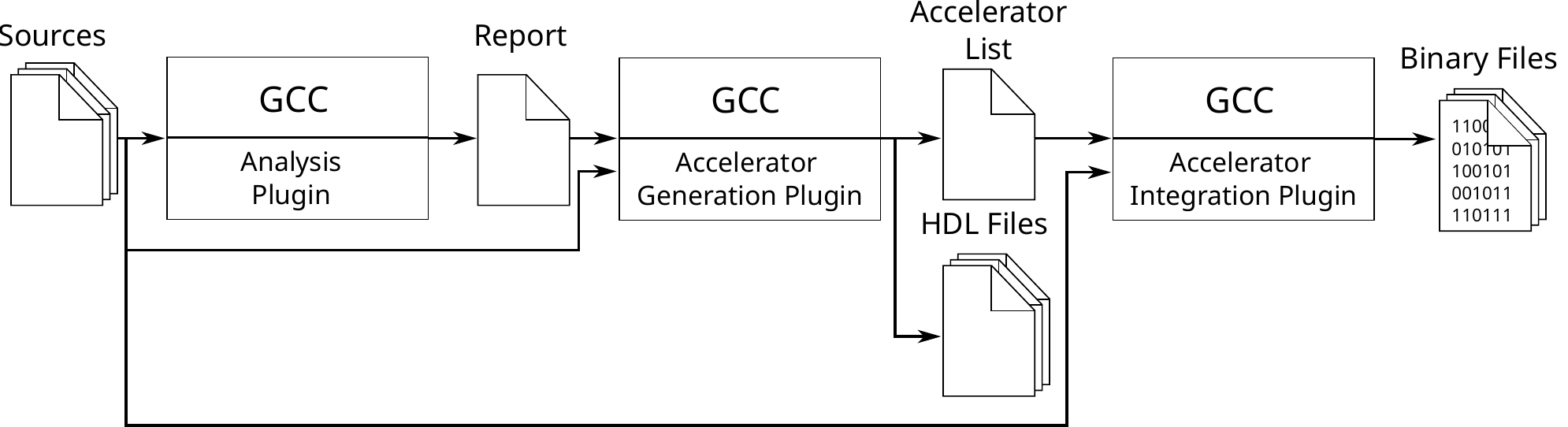}
\caption{Integration into GCC and workflow of PIRANHA. This process identifies segments of the code that can be accelerated using a hardware accelerator.
}
\label{fig:piranha}
\end{figure}

The working principle of PIRANHA is shown in Fig. \ref{fig:piranha}. The firmware for the LF-deck is processed through PIRANHA by means of three separate compiler plugins. The first collects information about the loops in the firmware. Loops that would benefit from parallelization or hardware acceleration are automatically identified and extracted. In a second call, the \texttt{Accelerator Generation} stage, each loop is separately optimized and scheduled. If the resulting schedule is faster than a pure software implementation, PIRANHA generates the corresponding hardware accelerator and modifies the original firmware in the third and final call through the \texttt{Accelerator Integration} plugin. Here, the accelerated loop is removed from the firmware and replaced with a call to the previously generated hardware accelerator. This process is transparent to the programmer, any part of the software that is not suitable for acceleration will be executed on the MicroBlaze as expected. PIRANHA takes advantage of the various optimization strategies and static analysis methods that the GCC provides to create fast and efficient hardware accelerators. If no conclusion about independence of operations is possible based on static analysis, run-time conditions can be generated \cite{loopParallelisation}. Likewise, bypasses and squashes are automatically applied to series of dependent memory accesses to further improve the accelerator throughput \cite{relaxedMemoryAccessScheduling}. For the neural network evaluation, the feed-forward calculation is accelerated for both the vector-matrix multiplication as well as for the activation function.  
\subsection{Fixed-Point Arithmetic}
\label{subsec:fixed-point}

The process of converting the weights of the DNN into integers involves using fixed-point arithmetic. This conversion is achieved by multiplying each weight by $2^n$ and then rounding down the result to get an integer value, where $n \in \mathbb{N}$ represents the amount of fractional bits. For DNNs with multiple Multilayer Perceptron (MLPs) we apply a uniform transformation for conversion, meaning the same $n$ value is used for all the MLPs within the DNN. 
To determine the optimal number of fractional bits, we randomly generate input samples for the DNN and run them through the network to get the output as a floating point number. The inputs are also converted into their fixed-point counterparts using the chosen $n$, and then they are fed into the fixed-point version of the DNN to obtain another set of outputs, this time represented as integers, which are then converted to floating point numbers by being multiplied with $2^{-n}$. By comparing the outputs from the floating-point and fixed-point version of the DNN, the maximum error for each $n$ can be assessed. This process is repeated with different samples for each $n$ value in ascending order, starting from $n=1$. We choose the $n$ that results in the smallest maximum difference between the floating-point and fixed-point representations.  

\section{MODEL}
In the following we are going to discuss the dynamic model of the quadrotor and the model of the DNN used in this work. For training, we use the simulator proposed in \cite{huang2023quadswarm}, which includes a dynamic and physical model of the quadrotor according to \cite{8967695}, along with a RL interface. We use the same reward function and hyper parameters described in \cite{batra2022decentralized}, and use the RL interface in \cite{huang2023quadswarm} to modify the deepsets architecture proposed in \cite{batra2022decentralized} to simplify the implementation on the LF-deck. 

\subsection{Quadrotor Dynamics}
\label{subsec:q_dynamics}
The quadrotor is modeled as a rigid body, where $\boldsymbol{x}$ denotes the position of its center of mass in an inertial reference frame. The rotation matrix from the quadrotor body frame to the inertial frame is denoted as $\boldsymbol{R}$. The equation of motion can be written as follows using the Newton-Euler equations
\begin{equation} \label{eq1}
\begin{split}
    \ddot{\boldsymbol{x}} &= \boldsymbol{g} + \frac{\boldsymbol{R} \boldsymbol{f}}{m} \\
    \dot{\boldsymbol{\omega}} &= \boldsymbol{I}^{-1}(\boldsymbol{\boldsymbol{\tau}} - \boldsymbol{\boldsymbol{\omega}} \times (\boldsymbol{I} \cdot \boldsymbol{\boldsymbol{\omega}})) \\
    \dot{\boldsymbol{R}} &= \boldsymbol{R} \boldsymbol{\boldsymbol{\omega_{\times}}},
\end{split}
\end{equation}
where $\boldsymbol{g}$ is the gravity vector, $m$ is the mass of the quadrotor, $\boldsymbol{f}$ is the total thrust vector in body frame, $\boldsymbol{\omega}$ is the angular velocity in the body frame and $\boldsymbol{I}$ is the inertia matrix. $\boldsymbol{\omega_{\times}}$ is the skew symmetric matrix of $\boldsymbol{\omega}$.

\subsection{Reinforcement Learning}
\label{subsec:rl}
In the following $t$ denotes the time step and $q$ the observing quadrotor. The self observation vector is defined as $\boldsymbol{o}^{q}_{t} = (\boldsymbol{p}_{t}^{q,j},\boldsymbol{v}_{t}^{q},\boldsymbol{r}_{t}^{q},\boldsymbol{\omega}_{t}^{q})$, where   $\boldsymbol{p}_{t}^{q,j} \in \mathbb{R}^{3}$ is the position of the quadrotor relative to its own target position, i.e.  $\boldsymbol{p}_{t}^{q,j} = \boldsymbol{p}_{t}^{q} - \boldsymbol{\hat{p}}_{t}^{j}$, where  $\boldsymbol{p}_{t}^{q}$ and $\boldsymbol{\hat{p}}_{t}^{j}$ are its own position and  target position respectively. $\boldsymbol{v}_{t}^{q} \in \mathbb{R}^{3}$ is the velocity of the quadrotor in the global world frame,  $\boldsymbol{r}_{t}^{q} \in \mathbb{R}^{9}$ is the rotation vector which we obtain by row-wise flattening  the rotation matrix $\boldsymbol{R}_{t}^{q} \in \mathbb{R}^{3 \times 3}$ (from the quadrotor body frame to the world frame)
, and $\boldsymbol{\omega}_{t}^{q}$ is the angular velocity of the quadrotor in body frame. The neighbor observations are $\boldsymbol{\hat{o}}^1_{t},..., \boldsymbol{\hat{o}}^{l}_{t},..., \boldsymbol{\hat{o}}^{k}_{t}$ with $k$ neighbors, where the $l$th vector is related to the $l$th nearest neighbor neighbor, and is denoted as $\boldsymbol{\hat{o}}^l_{t} = (\boldsymbol{p}_{t}^{q,l},\boldsymbol{v}_{t}^{q,l}) \in \mathbb{R}^{6}$, where $\boldsymbol{p}_{t}^{q,l}$ and $\boldsymbol{v}_{t}^{q,l}$ are the position and velocity of the $q$th quadrotor relative to the $k$th neighbor respectively, defined as $\boldsymbol{p}_{t}^{q,l} = \boldsymbol{p}_{t}^{q} - \boldsymbol{p}_{t}^{l}$ and $\boldsymbol{v}_{t}^{q,l} = \boldsymbol{v}_{t}^{q} - \boldsymbol{v}_{t}^{l}$, where $\boldsymbol{p}_{t}^{l}$ and $\boldsymbol{v}_{t}^{l}$ are the global position and velocity of the $l$th neighbor respectively. The output of the neural network is defined as $\boldsymbol{a} \in [-1,1]^4$, and we obtain the thrust vector by applying an affine transformation $\boldsymbol{\hat{f}} = \frac{1}{2}(\text{clip}(\boldsymbol{a},-1,1) + 1)$. 

\begin{figure}[t]
\centering
\includegraphics[scale=0.2]{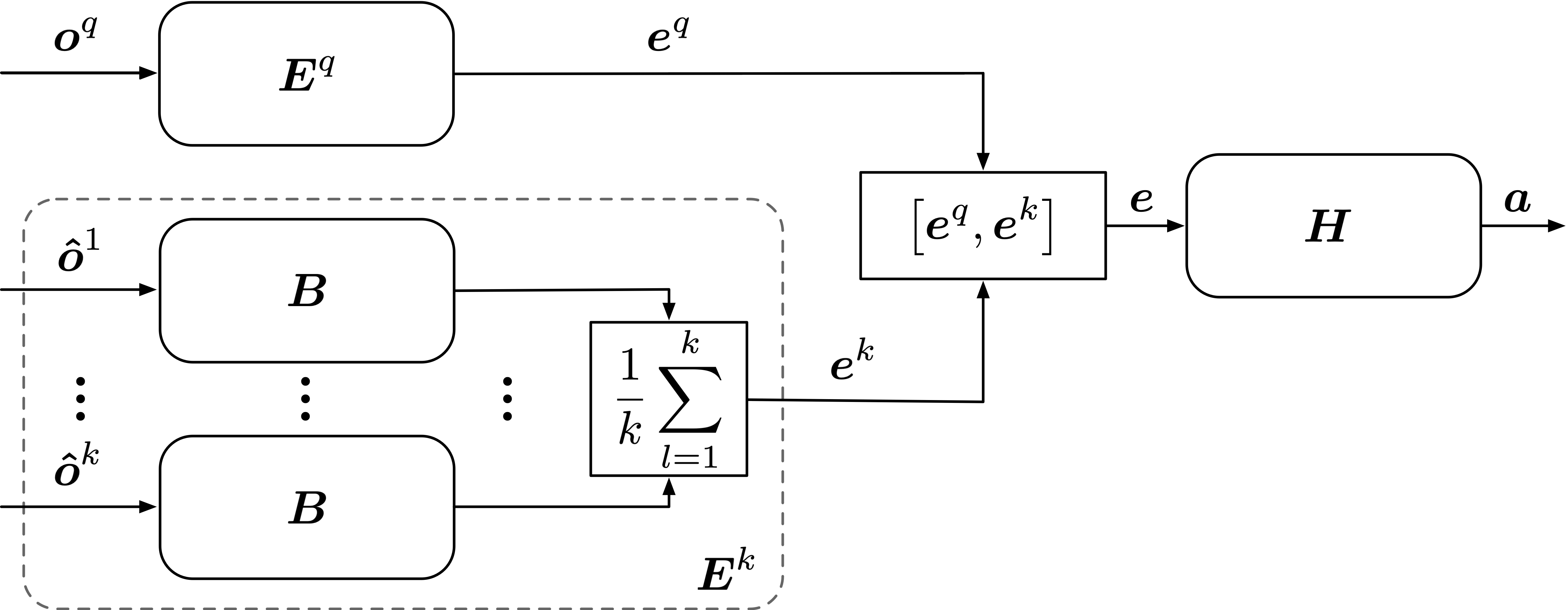}
\caption{The neural network structure implemented in the work. The network consists of the self encoder $\boldsymbol{E}^q$ and the neighbor encoder $\boldsymbol{E}^k$. The neighbor encoder consists of the MLP $\boldsymbol{B}$ and a mean operator for calculating the mean vector $\boldsymbol{e}^k$ out of the output vectors of $\boldsymbol{B}$.  The output of both encoders are concatenated, where $[ \boldsymbol{e}^q, \boldsymbol{e}^k ]$ denotes the concatenation, and processed in the output network $\boldsymbol{H}$.}
\label{fig:nn}
\end{figure}

The DNN model consists of the self encoder $\boldsymbol{E}^q$, which is an MLP, that processes the self observation, and the neighbor encoder $\boldsymbol{E}^k$ that processes the neighbor observations. The DNN model we use is permutation and scale invariant. For this, each neighbor vector is inserted to the MLP $\boldsymbol{B}$, and the mean vector is calculated out of the output vectors that are obtained from  $\boldsymbol{B}$, to obtain the output of the neighbor encoder $\boldsymbol{e}^{k}$. Finally the output of the self encoder $\boldsymbol{e}^{q}$ and the neighbor encoder $\boldsymbol{e}^{k}$ are concatenated and inserted to the MLP $\boldsymbol{H}$, to obtain the output of the neural network, denoted as $\boldsymbol{a}$. This structure is depicted in Figure \ref{fig:nn}. The self encoder consists of two hidden layers with $16$ neurons each and ReLU activation functions after each hidden layer. The neighbor encoder has two hidden layers and $8$ neurons and ReLU activation functions after each hidden layer. The MLP $\boldsymbol{H}$ has one hidden layer with $32$ neurons, and a ReLU activation function, where the output layer is without an activation function.

\section{EXPERIMENTS AND RESULTS}
\begin{figure}[h!]
\hspace{-.3cm}
 \includegraphics[scale=0.6]{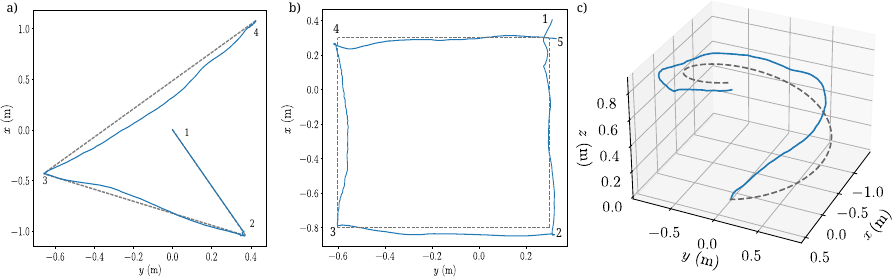}
\caption{Flight trajectories of the Crazyflie, where the gray dashed line and blue line are the desired flight trajectory and the real flight trajectory respectively. a) 4 setpoints in different direction. b) Rectangular flight trajectory tracking. c) Spiral flight trajectory tracking.}
\label{fig:traj}
\end{figure}

To validate our implementation we show flight trajectories of the Crazyflie that is equipped with the LF-deck.
We conduct 3 experiments in a flight area of $6.5 m \times 4.5 m \times 2.7 m$, where the results are shown in the figure \ref{fig:traj}. 

\textit{a) 4 Setpoints in different directions:} In this scenario, the Crazyflie starts at the position 1, and the setpoints 2, 3 and 4 are arbitrarily placed in different directions in order to show the ability of the learned policy to react to sudden changes in the direction of the target position. As we can see, the Crazyflie is able to arrive at the setpoints while maintaining a small distance to the desired target positions. The overall performance shows that Crazyflie is able to fly towards target positions at different directions with a small deviation from the desired trajectory path, which means that our implementation of the DNN on the FPGA was successful. 

\textit{b) Rectangular flight trajectory:} The Crazyflie flies along a rectangular path. It starts at point 1 and lands at the end at point 5. We can observe a deviation from the desired path, in particular between the setpoints 2 to 3 and 3 to 4. However, we also see that the Crazyflie is able to fly to its target position, demonstrating the functionality of the low level control policy of the neural network. 

\textit{c) Spiral flight trajectory:} The Crazyflie is placed at the center of the room, and performs a spiral flight trajectory. The Crazyflie flies to the desired target setpoints, where the last setpoint lies above the starting position at $z=0.8\text{m}$. Here we see a notable deviation from the desired trajectory, although the Crazyflie was able to fly according to a spiral trajectory. This performance shows the successful implementation of the DNN on the LF-deck. 

\section{CONCLUSIONS AND FUTURE WORK}
In this work we have presented the successful implementation of DNN on the LF expansion deck. The trained neural network is capable of low-level motor control while avoiding collisions among the quadrotors, where in this work we demonstrated only low level control on a real quadrotor. The implementation of computation expensive task on the LF-deck can be utilized to accelerate the computation and by this also to consume less energy. We validate our implementation by showing successful flight trajectories in different flight scenarios. To the best of our knowledge, this is the first implementation of the deepsets architecture on an FPGA extension deck for the Crazyflie quadrotors, using a cost-efficient positioning system that allows on-board position estimation. In the future, we plan to implement larger DNNs on the LF-deck as well as other network structures such as the attention model \cite{batra2022decentralized}. A detailed comparison between the LF-deck and the STM32F405 in terms of power consumption and execution time will be conducted. In addition, more LF-decks will be manufactured to demonstrate swarming behavior while maintaining collision-free flight among a fleet of quadrotors.